\documentclass{jfm}
\usepackage{graphicx}
\usepackage{epstopdf, epsfig}

\usepackage{float}

\usepackage{amsmath}
\usepackage[colorlinks, citecolor=blue]{hyperref}
\usepackage{siunitx}
\usepackage[usenames,dvipsnames]{xcolor}
\usepackage{setspace}

\usepackage{lipsum}

\newcommand*\red{\textcolor{red}}

\newcommand{\Ohn}{\mathit{Oh}}
\newcommand{\Ohc}{\mathit{Oh}_\mathit{c}}
\newcommand{\Oha}{\mathit{Oh}_\mathit{a}}

\newcommand{\Wen}{\mathit{We}}
\newcommand{\Ren}{\mathit{Re}}
\newcommand{\Bon}{\mathit{Bo}}
\newcommand{\Boc}{\mathit{Bo}_\mathit{c}}

\newcommand{\revRev}[1]{\textcolor{Black}{#1}}

\shorttitle{When does an impacting drop stop bouncing?}
\shortauthor{V. Sanjay, P. Chantelot, and D. Lohse}

\title{When does an impacting drop stop bouncing?}
\author{
	Vatsal Sanjay\aff{1}
	\corresp{\email{vatsalsanjay@gmail.com}},
	Pierre Chantelot\aff{1}
 	\corresp{\email{p.r.a.chantelot@utwente.nl}},
	\and Detlef Lohse{\aff{1}$^{,}$\aff{2}}
\corresp{\email{d.lohse@utwente.nl}}}

\affiliation{\aff{1}Physics of Fluids Group, Max Planck Center for Complex Fluid Dynamics, Department of Science and Technology, and J. M. Burgers Centre for Fluid Dynamics, University of Twente, P. O. Box 217, 7500 AE Enschede, The Netherlands
\aff{2}Max Planck Institute for Dynamics and Self-Organization, Am Fassberg 17, 37077 G\"{o}ttingen, Germany}

\begin{document}
\maketitle

\begin{abstract}
Non-wetting substrates allow impacting liquid drops to spread, recoil, and takeoff, provided they are not too heavy \citep{biance2006} or too viscous \citep{jha2020viscous}. 
In this article, using direct numerical simulations with the volume of fluid method, we investigate how viscous stresses and gravity \revRev{oppose} 
capillarity to inhibit drop rebound. Close to the bouncing to non-bouncing transition, we evidence that the initial spreading stage can be decoupled from the later retraction and takeoff, allowing to understand the rebound as a process converting the surface energy of the spread liquid into kinetic energy.
Drawing an analogy with coalescence induced jumping, we propose a criterion for the transition from the bouncing to the non-bouncing regime, namely by the condition $\Ohc + \Boc \sim 1$, where $\Ohc$ and $\Boc$ are the Ohnesorge number and Bond number at the transition, respectively. This criterion is in excellent agreement with the numerical results.
We also elucidate the mechanisms of bouncing inhibition in the heavy and viscous drops limiting regimes by calculating the energy budgets and relating them to the drop's shape and internal flow.
\end{abstract}


\section{Introduction}
\label{sec:Intro}
\revRev{Evidence of scientists' fascination for drop impacts can be traced back to the sketch of a water drop splashing onto a sheet of paper by Leonardo da Vinci in the margin of folio 33r in Codex Hammer/Leicester (1506 -- 1510) \citep{da1508notebooks}.}
In particular, the striking patterns created by drop fragmentation, at high impact velocity, have attracted attention \citep{rein1993phenomena, xu2005drop, yarin2006drop, villermaux2011drop, josserand2016drop, kim2020raindrop}. 
Lower velocity impacts, although they do not cause drops to shatter, also give rise to a rich variety of phenomena \citep{worthington1877xxviii, worthington1877second, chandra1991collision, thoroddsen2008high, yarin2006drop, josserand2016drop}.
The rebound of drops on non-wetting substrates may be one of the most fascinating of such interactions \citep{richard2000bouncing, richard2002contact, tsai2009drop, nair2014leidenfrost}.

Upon impact, the liquid first spreads \citep{Philippi2016, Gordillo2018} until it reaches its maximal extent \citep{Clanet2004,laan2014maximum,wildeman-2016-jfm,gordillo-2019-jfm}. 
It then recoils, following a Taylor-Culick type retraction parallel to the substrate \citep{taylor-1959-procrsoclonda, culick-1960-japplphys, bartolo2005retraction, pierson2020revisiting, deka-2020-prf, sanjay2022taylor}, and ultimately bounces off in an elongated shape perpendicular to the substrate \citep{richard2000bouncing, yarin2006drop, josserand2016drop}. 

Such rebounds abound in nature, as non-wetting surfaces provide plants and animals a natural way to keep dry \citep{neinhuis1997characterization, quere2008wetting}, and are relevant in many industrial processes such as inkjet printing \citep{lohse2022fundamental}. In some applications, it is pertinent that drops ricochet off the surface, such as self-cleaning \citep{blossey2003self}, keeping clothes dry \citep{liu2008hydrophobic}, and anti-fogging surfaces \citep{mouterde2017antifogging}. 
However, in most cases, bouncing must be suppressed. \revRev{For example, in cooling applications \citep{kim2007spray, shiri2017heat, jowkar2019rebounding} and pesticide spraying in agriculture \citep{bergeron2000controlling, he2021optimization, gorin2022universal}.} It is therefore natural to wonder when a drop stops bouncing.

So, when does the bouncing stop?
On the one hand, \citet{biance2006} found that heavy drops, \emph{i.e.}, drops larger than their gravito-capillary length $l_c = \sqrt{\gamma/\rho_dg}$, where $\gamma$ is the drop-air surface tension coefficient, $\rho_d$ is the density of the drop and $g$ is the acceleration due to gravity, cannot bounce.
On the other hand, \citet{jha2020viscous} showed that there exists a critical viscosity, two orders of magnitude higher than that of water, beyond which aqueous drops do not bounce either, irrespective of their impact velocity. So, gravity and viscosity counteract the bouncing.

In this paper, we investigate and quantify how exactly gravity and viscous stresses \revRev{compete} 
against capillarity to prevent drops from bouncing off non-wetting substrates, using direct numerical simulations. 
We focus on evidencing the mechanisms of bouncing inhibition, and exhibit a simple criterion delineating the bouncing to non-bouncing transition through an analogy with coalescence-induced drop jumping \citep{boreyko2009, liu2014numerical, farokhirad2015coalescence,  mouterde2017merging, lecointre2019ballistics}.

The paper is organized as follows: \S~\ref{sec:method} discusses the governing equations employed in this work. \S~\ref{sec:bouncingInhibition} explores the bouncing to non-bouncing transition and formulates a criterion for the inhibition of bouncing based on first principles followed by \S~\ref{sec:LimitingCases} which delves into the limiting cases of this criterion. The paper ends with conclusions and an outlook on future work in \S~\ref{sec:Conclusion}. 

\section{Governing equations}
\label{sec:method}
\begin{figure}
	\centering
	\includegraphics[width=\textwidth]{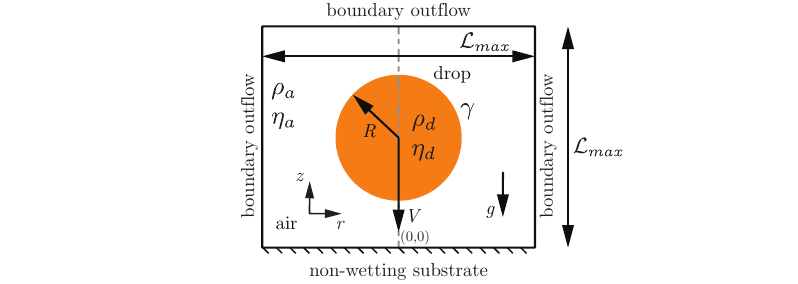}
	\caption{Axi-symmetric computational domain used to study the impact of a drop with radius $R$ and velocity $V$ on an ideal non-wetting substrate. The subscripts $d$ and $a$ denote the drop and air, respectively, to distinguish their material properties, the density $\rho$ and viscosity $\eta$. The drop-air surface tension coefficient is $\gamma$ and $g$ denotes the acceleration of gravity. The gray dashed-dotted line represents the axis of symmetry, $r = 0$. Boundary outflow is applied at the top and side boundaries (tangential stresses, normal velocity gradient, and ambient pressure are set to zero). The domain boundaries are far enough not to influence the drop impact process ($\mathcal{L}_{\text{max}} \gg R$, $\mathcal{L}_{\text{max}} = 8R$ in the worst case).}
	\label{fig:schematic}
\end{figure}

We employ direct numerical simulations to study the drop impact process in an axisymmetric setting (figure~\ref{fig:schematic}), using the free software program \emph{Basilisk C} \citep{popinet-basilisk} that employs the geometric volume of fluid (VoF) method for interface reconstruction \citep{popinet2009accurate}. For an incompressible flow, the mass conservation requires the velocity field to be divergence-free (tildes denote dimensionless quantities throughout this manuscript),

\begin{align}
	\boldsymbol{\tilde{\nabla}\cdot \tilde{v}} = 0,
\end{align}

\noindent where we non-dimensionalise the velocity field with the inertio-capillary velocity $V_\gamma = \sqrt{\gamma/(\rho_d R)}$.
We further non-dimensionalise all lengths with the drop radius $R$ (figure~\ref{fig:schematic}), time with the inertio-capillary timescale $\tau = \sqrt{\rho_dR^3/\gamma} = R/V_\gamma$, and pressure with the capillary pressure, $p_\gamma = \gamma/R$, to write the momentum equation as

\begin{align}
	\label{eq::NS}
	\frac{\partial \boldsymbol{\tilde{v}}}{\partial \tilde{t}} + \boldsymbol{\nabla \cdot}\left(\boldsymbol{\tilde{v}}\boldsymbol{\tilde{v}}\right) = \frac{1}{\tilde{\rho}}\left(-\boldsymbol{\tilde{\nabla}} \tilde{p}^{\prime} + \boldsymbol{\tilde{\nabla}\cdot}\left(2\Ohn\boldsymbol{\tilde{\mathcal{D}}}\right)  + \boldsymbol{\tilde{f}}\right),
\end{align}

\noindent where the deformation tensor $\boldsymbol{\mathcal{D}}$ is the symmetric part of the velocity gradient tensor $\left(= \left(\nabla\boldsymbol{v} + \left(\nabla\boldsymbol{v}\right)^{\text{T}}\right)/2\right)$.
The Ohnesorge number $\Ohn$ (the ratio of inertio-capillary to inertio-viscous time scales) and the dimensionless density $\tilde{\rho}$ are written using the one-fluid approximation \citep{prosperetti2009computational, tryggvason2011direct} as 

\begin{align}
	\label{eq::Oh}
	\Ohn &= \Psi\Ohn_d + \left(1-\Psi\right)\Oha,\\
	\label{eq::density}
	\tilde{\rho} &= \Psi + \left(1-\Psi\right)\frac{\rho_a}{\rho_d},
\end{align}

\noindent where $\Psi$ is the VoF tracer ($= 1$ for the drop and $0$ otherwise), and $\rho_a/\rho_d$ is the air--drop density ratio. Here

\begin{align}
	\Ohn_d = \frac{\eta_d}{\sqrt{\rho_d\gamma R}} \quad \text{and} \quad \Oha = \frac{\eta_a}{\sqrt{\rho_d\gamma R}}
\end{align}

\noindent are the Ohnesorge numbers based on the viscosities of the drop liquid and of air, respectively. To minimize the influence of the surrounding medium, we keep $\rho_a/\rho_d$ and $\Oha$ fixed at $10^{-3}$ and $10^{-5}$, respectively. For a lean notation, we will use 
$\Ohn$ instead of $\Ohn_d$ in the remainder of the text. 

Lastly, $\tilde{p}^{\prime}$ denotes the reduced pressure field, $\tilde{p}' = \tilde{p}\,+\,\Bon\tilde{\rho}\tilde{z}$, where, $\tilde{p}$ and $\Bon\tilde{\rho}\tilde{z}$ represent the mechanical and the hydrostatic pressures, respectively. 
Here, $\tilde{z}$ is the distance away from the non-wetting substrate (see figure~\ref{fig:schematic}) and the Bond number $\Bon$ compares gravity to the surface tension force,

\begin{align}
	\Bon = \frac{\rho_d g R^2}{\gamma}.
\end{align}

\noindent Using this reduced pressure approach ensures an exact hydrostatic balance as described in \citet{popinet2018numerical, basiliskPopinet3}. This formulation requires an additional singular body force at the interface such that $\boldsymbol{\tilde{f}}$ takes the form \citep{brackbill1992continuum}

\begin{align}
	\boldsymbol{\tilde{f}} \approx \left(\tilde{\kappa} + \Bon\left(1 -\frac{\rho_a}{\rho_d}\right)\tilde{z}\right)\boldsymbol{\tilde{\nabla}}\Psi,
\end{align}

\noindent where the first and second terms on the right-hand side are the local capillary and hydrostatic pressure jumps across the interface, respectively with $\tilde{\kappa}$ the interfacial curvature calculated using the height-function approach \citep{popinet2009accurate}. 

Figure~\ref{fig:schematic} shows the axi-symmetric computational domain where we solve the equations discussed above. A no-slip and non-penetrable boundary condition is applied on the substrate along with a zero normal pressure gradient. Here, we also impose $\Psi = 0$ to maintain a thin air layer between the drop and the substrate to model an ideal non-wetting substrate \citep[for a detailed discussion about this method, readers are referred to][]{VatsalThesis}. Physically, it implies that the minimum thickness of this air layer is $\Delta/2$, where $\Delta$ is the minimum grid size, throughout the simulation duration. We use \emph{Basilisk C}'s \citep{popinet-basilisk} adaptive mesh refinement capabilities to finely resolve regions of high velocity gradients and at the drop-air interface. We undertook a mesh independence study to ensure that the results are independent of this mesh resolution and use a minimum grid size $\Delta = R/1024$ for this study. Initially, we assume that the drop is spherical and that it impacts with a dimensionless velocity, $\tilde{V} = V/V_\gamma = \sqrt{\Wen}$, where the impact Weber number 

\begin{align}
	\Wen = \frac{\rho_d R V^2}{\gamma}
\end{align}

\noindent is the ratio of the inertial pressure during impact to the capillary pressure. We refer the readers to \cite{popinet2009accurate, popinet-2015-jcp, popinet-basilisk, zhang2022impact, basiliskVatsalViscousBouncing, VatsalThesis} for details of the computational method employed in this work.

\section{Bouncing inhibition}
\label{sec:bouncingInhibition}

\begin{figure}
	\centering
	\includegraphics[width=\textwidth]{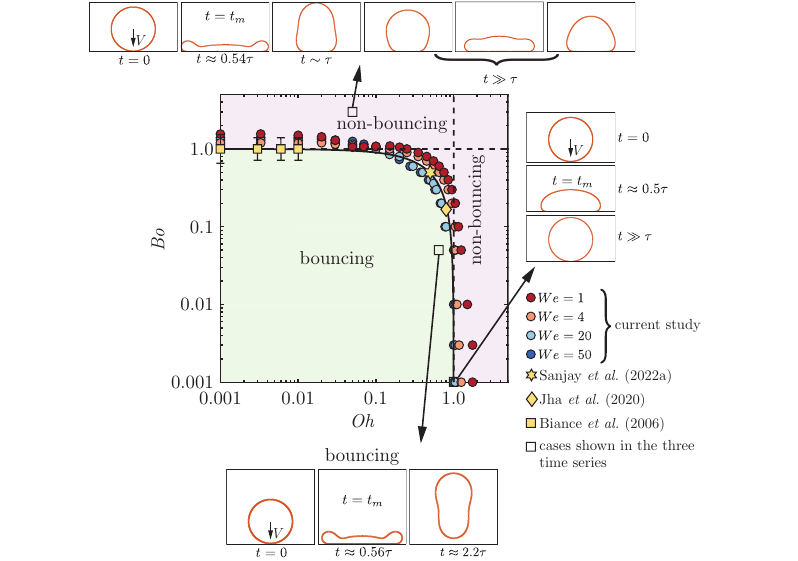}
	\caption{Regime map in terms of the Bond number $\Bon = \rho_dgR^2/\gamma$ and the drop Ohnesorge number $\Ohn = \eta_d/\sqrt{\rho_d\gamma R}$, distinguishing the bouncing and non-bouncing regimes. The data points represent the transition between the bouncing and non-bouncing regimes at different Weber numbers $\Wen$. The three series of insets illustrate typical cases in these regimes, namely $(\Wen, \Ohn, \Bon) = $ $(16, 0.05, 3)$ for the upper, $(16, 1, 0.001)$ for the right, and $(16, 0.75, 0.05)$ for the bottom series of images, respectively. The solid black line delineates the prediction of this transition (equation~\eqref{eq:MainEquation}). Lastly, the black dashed vertical and horizontal lines mark the two \revRev{asymptotes, i.e., the viscous limiting case}, $\Ohc = 1$ and \revRev{the weight limiting case,} $\Boc = 1$, respectively. See also the supplementary movie~\red{SM1}.}
	\label{fig:RegimeMap}
\end{figure}
\begin{figure}
	\centering
	\includegraphics[width=\textwidth]{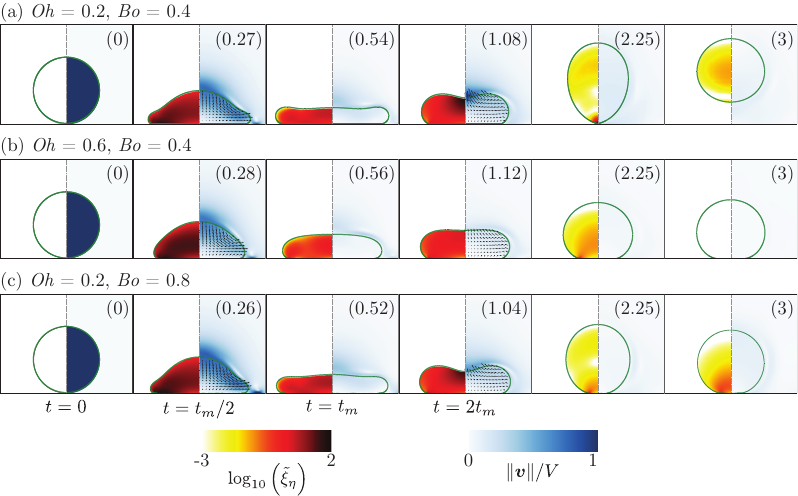}
	\caption{Direct numerical simulations snapshots illustrating the drop impact dynamics for $\left(\Ohn, \Bon\right)$ = $\left(0.2, 0.4\right)$ (a), $\left(\Ohn, \Bon\right)$ = $\left(0.6, 0.4\right)$ (b), and $\left(\Ohn, \Bon\right)$ = $\left(0.2, 0.8\right)$ (c). The left hand side of each numerical snapshot shows the dimensionless viscous dissipation function $\tilde{\xi}_\eta = 2\Ohn\left(\boldsymbol{\tilde{\mathcal{D}}:\tilde{\mathcal{D}}}\right)$ on a $\log_{10}$ scale to identify regions of maximum dissipation (black). The right hand side shows the magnitude of the velocity field normalized by the initial impact velocity, $V$. The black velocity vectors are plotted in the drop's \revRev{centre} of mass reference frame to evidence the internal flow. \revRev{The numbers inside the bracket at right-top corner of each snapshot represent the dimensionless time, $t/\tau$.} For all cases, the impact Weber number is $\Wen = 20$. See also the supplementary movie~\red{SM2}.}
	\label{fig:Phenomenology}
\end{figure}

We investigate the behavior of drops impacting on non-wetting substrates by exploring the influence of the following dimensionless parameters: the Weber number $\Wen = \rho R V^2/\gamma$, the Bond number $\Bon = \rho_dgR^2/\gamma $, and the drop Ohnesorge number $\Ohn = \eta_d/\sqrt{\rho_d\gamma R}$. 
\revRev{We restrict ourselves to impacts with $\Wen \ge 1$ and do not discuss the bouncing to non-bouncing transition observed for $\Wen \ll 1$ (gentle deposition) \citep{richard2000bouncing,molavcek2012quasi,planchette2012transition}}
In Figure~\ref{fig:RegimeMap}, we evidence the bouncing to non--bouncing transition in the parameter space spanned by the Ohnesorge and Bond numbers for several fixed Weber numbers. 
We extract three key pieces of information from this regime map: 
\begin{itemize}
	\item The Weber number has a small influence on the transition between the bouncing and non--bouncing regime in the range probed in this study, $We = 1 - 50$, similarly as reported by \revRev{\citet{jha2020viscous,antonini2016contactless}} for the bouncing inhibition of viscous drops (see also appendix~\ref{app:Weber}). 
	\item We recover the two limiting cases of non--bouncing (see insets of figure~\ref{fig:RegimeMap}): drops smaller than their visco-capillary length, ({\it i.e.}, $R < \eta_d^2/\rho_d\gamma$, giving $\Ohn > 1$) stop bouncing due to viscous dissipation \citep{jha2020viscous}, while those larger than their gravito-capillary length, ({\it i.e.}, $R > \sqrt{\gamma/\rho_d g}$, giving $\Bon > 1$) cannot bounce due to their weight \citep{biance2006}. We will elaborate on the mechanisms of rebound inhibition in  these two non--bouncing regimes in \S~\ref{sec:LimitingCases}.
	\item Experiments performed with millimetre--sized drops of water or silicone oil do not lie on either asymptote \revRev{\citep{jha2020viscous,sanjay2022drop}}, suggesting that both the effect of viscosity and gravity need to be taken into account to predict the bouncing to non--bouncing transition.
\end{itemize}

In this section, we focus on situations where bouncing is prevented by both viscous and gravitational effects ({\it i.e.}, $\Bon < 1$ and $\Ohn < 1$).
Figure~\ref{fig:Phenomenology} shows snapshots illustrating three representative cases lying in this region of the parameter space for $\Wen = 20$. Each snapshot displays three pieces of information: the position of the liquid--air interface, the dimensionless rate of viscous dissipation per unit volume ({\it i.e.}, the dimensionless viscous dissipation function, left panel), and the magnitude of the velocity field normalized with the initial impact velocity (right panel). For $\Ohn = 0.2$ and $\Bon = 0.4$ (figure~\ref{fig:Phenomenology}a), the drop undergoes typical rebound dynamics. The liquid first spreads radially up to $t = t_m$, when the maximum extent is reached \citep{Clanet2004, eggers2010drop, laan2014maximum, wildeman-2016-jfm}. This stage is followed by liquid retraction \citep{bartolo2005retraction}, parallel to the substrate, until the drop contracts ($t = 2t_m$) and the motion becomes vertical \citep{chantelot2018rebonds, zhang2022impact}. Finally, the drop leaves the substrate at $t = 2.25\tau$ \citep{richard2000bouncing, richard2002contact}. 

Surprisingly, increasing $\Ohn$ to $0.6$, below the critical value reported by \citet{jha2020viscous}, while keeping $\Bon = 0.4$ (figure~\ref{fig:Phenomenology}b), prevents the rebound. The motion is damped before the drop can bounce off the substrate.
Similarly, increasing $\Bon$ to 0.8, below the critical value reported by \citet{biance2006}, while fixing $\Ohn = 0.2$ (figure~\ref{fig:Phenomenology}c), also inhibits bouncing.  Yet, the deposited liquid undergoes multiple oscillation cycles on the substrate before coming to rest (see the last snapshot $t = 3\tau$).

In all three cases, the impact dynamics and flow in the drop are qualitatively similar until the maximum extent is reached at $t=t_m$. 
At this instant, the absence of internal flow suggests that the initial kinetic energy has either been converted into surface energy or lost to viscous dissipation, which occurs throughout the drop volume owing to $\Ohn \sim \mathcal{O}\left(0.1\right)$ \citep{eggers2010drop}.
Close to the bouncing to non--bouncing transition, the rebound can thus be understood as a process which converts an initial surface energy into kinetic energy, disentangling the later stages of the rebound from the initial impact dynamics.

This observation prompts us to introduce an analogy with coalescence-induced jumping, in which an excess surface energy, gained during coalescence, is converted into upward motion of the liquid \citep{boreyko2009, liu2014numerical, farokhirad2015coalescence}. The spread drop, at rest at $t=t_m$, reduces its surface area through a Taylor-Culick type retraction, converting excess surface energy into kinetic energy.
The capillary force driving this radially inwards flow is

\begin{align}
	\label{eq:drivingCapillary}
	F_\gamma \sim \gamma R.
\end{align}

\noindent \revRev{The velocity $v$ associated to this Taylor-Culick type retraction scales as $v \sim \sqrt{\gamma/\left(\rho_d e\right)}$, where $e$ is the typical thickness of the spread liquid, which can be rearranged as $v \sim V_\gamma \sqrt{R/e}$ upon introducing the inertio-capillary velocity \citep{bartolo2005retraction,chantelot2018rebonds}.
At the bouncing to non-bouncing transition, in the presence of both viscous and gravitational effects (see figure 2), we make the hypothesis that $e \sim R$, as no pronounced central film forms during spreading (see figure \ref{fig:Phenomenology} at $t=t_m$), implying that the inertio-capillary velocity is the relevant velocity scale, \emph{i.e.} $v \sim V_\gamma$.}
Similarly as in coalescence--induced jumping of two identical drops, a dissipative force $F_\eta \sim \Omega\eta_d\nabla^2v$, where $\Omega$ is the volume of the drop and $v$ is a typical radial flow velocity, opposes the capillarity driven flow \citep{mouterde2017merging, lecointre2019ballistics}.
Taking \revRev{$v \sim V_\gamma$ as explained above}, the resistive viscous force \revRev{then} scales as

\begin{align}
	\label{eq:resistVisc}
	F_\eta \sim \eta_d V_\gamma R,
\end{align}

\noindent and the effective momentum converging in the radial direction is

\begin{align}
	\label{eq:flowChange}
	P_r \sim \int \left(F_\gamma - F_\eta\right) \mathrm{d}t.
\end{align}

\noindent The asymmetry \revRev{originating} from the presence of the substrate enables the conversion of the radially inward momentum to the upwards direction (figure~\ref{fig:Phenomenology}, $t = 2t_m$). Following \citet{mouterde2017merging, lecointre2019ballistics}, we assume that the vertical momentum scales with the radial one, {\it i.e.}, $P_v \sim P_r$, allowing us to determine a criterion for the bouncing transition by balancing the rate of change of vertical momentum with the drop's weight $F_g$

\begin{align}
	\label{eq:competeGravity1}
	\frac{dP_v}{dt} = F_g \sim \rho_d R^3 g.
\end{align}

\noindent Using equations~\eqref{eq:drivingCapillary} --~\eqref{eq:flowChange}, we obtain 

\begin{align}
	\label{eq:competeGravity3}
	\gamma R - \eta_d V_\gamma R \sim \rho_dR^3g.
\end{align}

\noindent Lastly, substituting $V_\gamma = \sqrt{\gamma/\rho_dR}$, and rearranging, we arrive at a criterion to determine the bouncing to non-bouncing transition as 

\begin{align}
	\label{eq:MainEquation}
	\Ohc + \Boc \sim 1,
\end{align}

\noindent where the subscript $c$ stands for \lq critical\rq.
Equation \eqref{eq:MainEquation}, which is independent of the impact Weber number $\Wen$, is the main result of the manuscript.
 
We test the criterion \eqref{eq:MainEquation} for the bouncing to floating transition against data extracted from our direct numerical simulations and experiments from \citet{biance2006, jha2020viscous, sanjay2022drop}. 
In figure~\ref{fig:RegimeMap}, the solid black line, representing equation~\eqref{eq:MainEquation} with prefactor $1$, is in excellent quantitative agreement with the data when viscous and gravitational effects inhibit bouncing, as well as in the two limiting regimes, $\Ohc \sim 1$ for $\Bon \ll 1$ \citep{jha2020viscous}, and $\Boc \sim 1$ for $\Ohn \ll 1$ \citep{biance2006} (black dotted lines).

In the next section, we focus on evidencing the physical mechanisms leading to bouncing suppression in each of the two limiting cases.
\revRev{But before, we note that the data for different Weber numbers do not exactly collapse on the prediction of equation \eqref{eq:MainEquation} suggesting that the critical Ohnesorge and Bond numbers vary weakly with the Weber number, and hinting at the limitations of our hypothesis to choose $V_\gamma$ as the velocity scale, and to neglect the influence of $\Wen$ on the retraction velocity.}

\section{Limiting cases}\label{sec:LimitingCases}
\subsection{How does a viscous drop stop bouncing?}\label{sec:LimitingCases:Oh}
\begin{figure}
	\centering
	\includegraphics[width=\textwidth]{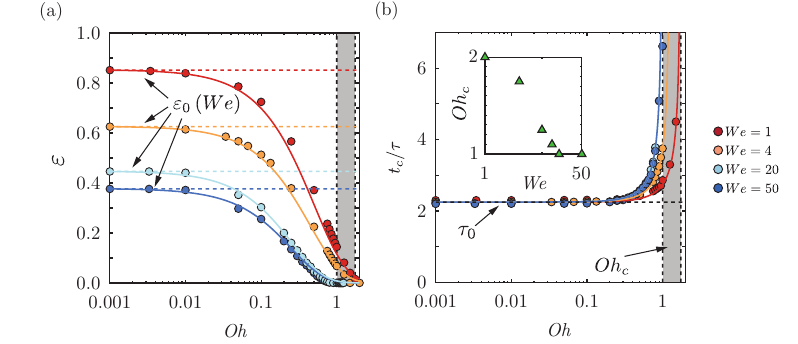}
	\caption{\revRev{Variation of (a) the restitution coefficient $\varepsilon$, and (b) the contact time $t_c$, normalized by the inertio-capillary timescale $\tau = \sqrt{\rho_dR^3/\gamma}$,} with the drop Ohnesorge number $\Ohn$ for $\Bon = 0$ at different Weber numbers $\Wen$. 
	In both panels, the solid lines represent the predictions of the spring-mass-damper system of \citet{jha2020viscous} (contact time, equation~\ref{eqn:JhaEtAl_time}, and restitution coefficient, equation~\ref{eqn:JhaEtAl_epsilon}). 
	The horizontal dashed lines represent the contact time and restitution coefficient values in the $\Ohn \ll 1$ limit, in which $\tau_0 = 2.25\tau$, independent of $\Wen$, while $\varepsilon_0\left(\Wen\right) = \varepsilon\left(\Wen, \Ohn \to 0, \Bon = 0\right)$ (equation~\eqref{Eqn:Epsilon0Def}) depends on $\Wen$.
	Lastly, the black vertical lines and the gray shaded regions mark the critical Ohnesorge number $\Ohc \sim \mathcal{O}\left(1\right)$ beyond which drops do not bounce.}
	\label{fig:Ohlim}
\end{figure}
\begin{figure}
	\centering
	\includegraphics[width=\textwidth]{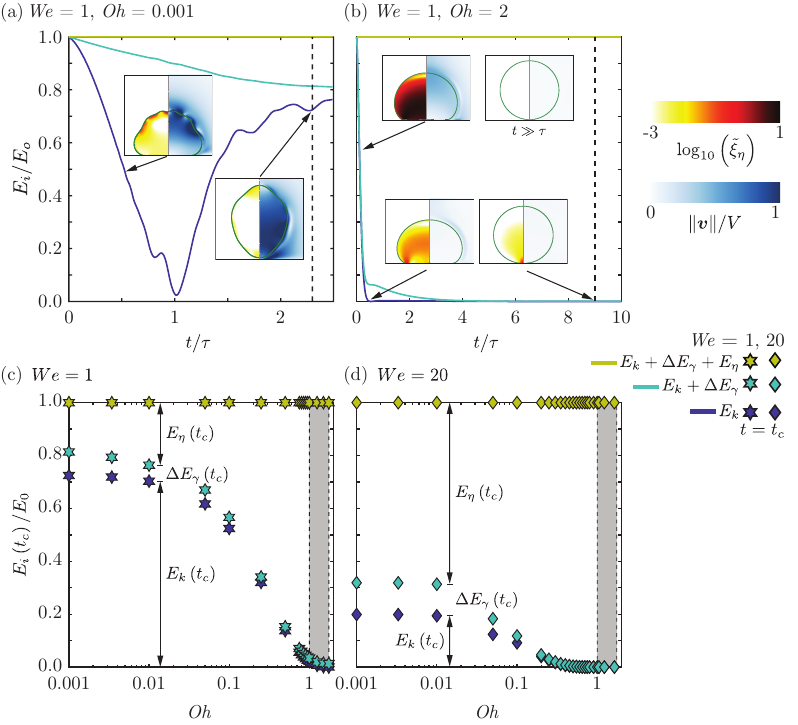}
	\caption{Energy budgets for drop impacts with $\Bon = 0$ and $\Wen = 1$ for $\Ohn = 0.001$ (a) and  $\Ohn = 2$ (b). $E_k$ and $E_\eta$ represent the kinetic energy and viscous dissipation, respectively. $\Delta E_\gamma$ denotes the change in surface energy with its zero set at $t = 0$. 
	The numerical snapshots in the insets illustrate the drop morphologies and the anatomy of the flow with a color code identical to that of figure \ref{fig:Phenomenology}.
	The black dotted lines in panels (a) and (b) mark the instant $t_c$ when the drop takes off and when the normal contact force between the drop and the substrate is minimum, respectively. 
	(c,d) Energy distributions at $t = t_c$ for $\Wen = 1$ (c) and $\Wen = 20$ (d) as function of $\Ohn$. The black vertical lines and the gray shaded regions mark the critical Ohnesorge number $\Ohc \sim \mathcal{O}\left(1\right)$ beyond which drops do not bounce. See also supplementary movie~\red{SM3}.}
	\label{fig:OhlimDescription}
\end{figure}

We first investigate how viscous drops, much smaller than their gravito-capillary length, {\it i.e.}, with $\Bon \ll 1$, stop bouncing.
We study this regime, in which the transition criterion \eqref{eq:MainEquation} reduces to $\Ohc \sim 1$, by setting $\Bon$ to $0$ \revRev{(i.e., by assuming that capillarity dominates over gravity)} and by systematically varying the drop Ohnesorge number, $\Ohn$.
We characterize the rebound behavior by measuring the apparent contact time $t_c$ between the drop and the substrate and the coefficient of restitution $\varepsilon$, that we define as $\varepsilon = v_\text{cm}(t_c)/V$, where $v_\text{cm}(t_c)$ is the \revRev{centre} of mass velocity at take-off.
The procedure used to extract $t_c$ and $\varepsilon$ from the DNS is detailed in appendix~\ref{app:restitution in simulations}.

In figure~\ref{fig:Ohlim}, we plot the coefficient of restitution $\varepsilon$ and the normalized contact time $t_c/\tau$ as a function of $\Ohn$ for Weber numbers ranging from 1 to 50.
The effect of $\Ohn$ on $\varepsilon$ and $t_c$ is markedly different.
On the one hand, the coefficient of restitution monotonically decreases from its low $\Ohn$, Weber dependent value

\begin{align}
	\varepsilon_0\left(\Wen\right) = \varepsilon\left(\Wen, \Ohn \to 0, \Bon = 0\right)
	\label{Eqn:Epsilon0Def}
\end{align}

\noindent with increasing $\Ohn$, until a critical Ohnesorge number of order one, $\Ohc$, marking the end of the bouncing regime is reached.
On the other hand, increasing $\Ohn$ by over two orders of magnitude hardly affects $t_c$. It keeps its Weber independent value $\tau_0 = 2.25\tau$, expected from the inertio--capillary scaling in the low $\Ohn$ limit \citep{wachters1966heat,richard2002contact}, until $t_c$ diverges as $\Ohn$ tends towards $\Ohc$.

Figure~\ref{fig:Ohlim} also highlights that $\Ohc$ varies weakly with $\Wen$ ($\Ohc = 1.75, 1.5, 1, 1$ at $\Wen = 1, 4, 20, 50$, respectively, see inset of figure~\ref{fig:Ohlim}\revRev{b}) as evidenced by the narrow gray shaded region, and in agreement with the limit predicted from equation~\eqref{eq:MainEquation}.
Varying $\Wen$ mainly affects the low $\Ohn$ restitution limit $\varepsilon_0\left(\Wen\right)$, which we elaborate on in appendix~\ref{app:Weber}.
We stress that the weak variation of the coefficient of restitution in the shaded region, where $\varepsilon<0.1$, could go unnoticed in typical side view experiments.
Indeed, $\varepsilon = 0.1$ corresponds to a \revRev{centre} of mass rebound height of $0.01$ times the initial impact height, that is $10\,\si{\micro\meter}$ for $\Wen = 1$.

We now seek to understand the evolution of \revRev{the restitution coefficient} $\varepsilon$ with $\Ohn$ by quantifying the overall energy budget during an impact event.
In the $\Bon = 0$ limit, the energy balance reads

\begin{align}
	\label{eqn:OhEnergyBalance}
	\tilde{E}_0 = \tilde{E}_k(\tilde{t}) + \Delta\tilde{E}_\gamma(\tilde{t}) + \tilde{E}_\eta(\tilde{t}).
\end{align}

\noindent where each energy component is normalized using the capillary energy scale $\gamma R^2$, $E_0$ denotes the drop's initial kinetic energy, ($\tilde{E}_0 = E_0/(\gamma R^2) = (2\pi/3)\Wen$), $E_k(t)$ and $E_\gamma(t)$ are the drop's time dependent kinetic and surface energies, with $\Delta E_\gamma(t) = E_\gamma(t) - E_\gamma(t = 0)$, and $E_\eta(t)$ is the viscous dissipation until time $t$. Readers are referred to \citet{landau2013course, wildeman-2016-jfm, ramirezsoto-2020-sciadv, sanjay2022taylor, VatsalThesis} for details of energy budget calculations. 

Figure \ref{fig:OhlimDescription}(a) evidences the time evolution of the energy balance contributions for an impact with $\Wen=1$ and $\Ohn=0.001$.
The drop's initial kinetic energy $E_0$ is transferred into surface energy until the liquid reaches its maximal extent at \revRev{$t = t_m$  \citep[note that for $We = 1$, $t_m \approx \tau$, see][]{zhang2022impact}}. At this instant, the energy available to the drop is almost exclusively stored in the form of excess surface energy, as hypothesized in our analogy with coalescence-induced jumping.
As the drop retracts, surface energy is converted back into kinetic energy and, at take-off, the drop recovers a large proportion of its initial kinetic energy, $E_k(t_c) \approx 0.75E_0$.
Energy dissipation throughout the rebound, $E_\eta(t_c)$, and the non-spherical drop shape at take-off, storing excess surface energy $\Delta E_\gamma(t_c)$, hamper the recovery of the initial kinetic energy.
Even in the low $\Wen$ and \revRev{low} $\Ohn$ case at hand, where dissipation is restricted to the boundary layer at the drop-air interface and happens due to the propagation of capillary waves \cite[see the insets of figure~\ref{fig:OhlimDescription}(a) and][]{renardy2003pyramidal, zhang2022impact}, viscous stresses dissipate $20$\% of the initial energy during the rebound. 

Increasing the drop Ohnesorge number to $\Ohn = 2$ does not affect the energy transfer dynamics (figure \ref{fig:OhlimDescription}b), but it enhances viscous dissipation, which now takes place in the whole liquid volume \cite[see the insets of figure~\ref{fig:OhlimDescription}(b) and][]{eggers2010drop}.
Beyond the critical Ohnesorge number $\Ohc$, the initial kinetic energy is dissipated before the drop can rebound off the substrate.
The drop impact process becomes over-damped and, in this small Bond number limit, the drop slowly relaxes back to its sessile spherical shape (figure~\ref{fig:OhlimDescription}b).

Figures \ref{fig:OhlimDescription}(c) and (d) summarize the distribution of energy at take-off as a function of $\Ohn$ for $\Wen = 1$ and $\Wen = 20$, respectively.
For $\Ohn < 0.01$, the overall energy budget is not affected by a change in drop Ohnesorge number, extending the validity domain of the so-called inviscid drop limit \citep{richard2000bouncing}.  
Strikingly, the independence of $E_\eta(t_c)$, and thus of $\varepsilon$, with $\Ohn$ in this limit does not imply that viscous dissipation is negligible. 
Indeed, (i) the dissipated energy accounts for more than two thirds of the total energy kinetic energy loss during impact at $\Wen = 1$, where the restitution is maximal, and (ii) the increase of viscous dissipation is mainly responsible for the decrease of $\varepsilon$ with $\Wen$. The dissipated energy $E_\eta(t_c)$ accounts for $20\%$ and $70\%$ of $E_0$ for $\Wen = 1$ and $\Wen = 20$, respectively, contradicting the inviscid nature of this regime. The transfer of the initial kinetic energy into surface energy $\Delta E_\gamma(t_c)$ at take-off, that is the rebound of the liquid in a non-spherical shape, while accounting for one third the total energy loss during impact at $\Wen = 1$, cannot explain alone the significantly lower than one value of the coefficient of restitution. 

The presence of a finite energy dissipation in the limit $\Ohn \to 0$ is reminiscent of the dissipative anomaly 
in fully developed turbulence, expressing that even in the limit of vanishing viscosity (i.e., diverging Reynolds number $\Ren \to \infty$), the energy dissipation rate remains finite \citep{onsager1949statistical, eyink1994energy, kolmogorov1941local, dubrulle2019beyond, eggers2018role}.
The dissipative anomaly 
reflects in the finite drag experienced by solid bodies at diverging 
Reynolds numbers, through the creation of boundary layers \citep{prandtl1904}, somewhat 
similar to the localization of viscous dissipation at the liquid-air interface during drop impact \revRev{\citep[see the inset of figure~\ref{fig:OhlimDescription}a and][]{Philippi2016}}.

For larger Ohnesorge numbers, the dissipated energy $E_\eta(t_c)$ increases with $\Ohn$, \revRev{reflecting} that viscous dissipation is responsible for the loss of the rebound elasticity. 
Interestingly, increasing $\Ohn$ also reduces the drop deformation at take-off, \revRev{decreasing the fraction of energy stored as surface energy $\Delta E_\gamma(t_c)$. Consequently, energy which is not lost to viscous dissipation is mainly converted back into the kinetic energy of the drop} leading to a more efficient recovery of the initial kinetic energy (figures \ref{fig:OhlimDescription}c,d).

We further rationalize our observations by comparing our simulation results to the predictions of \citet{jha2020viscous} that extend the liquid spring analogy to viscous drops.
This minimal model, that has been shown to successfully capture the variation of $t_c$ and $\varepsilon$ with $\Ohn$, gives the time of apparent contact as

\revRev{\begin{align}
	\label{eqn:JhaEtAl_time}
	t_c\left(\Wen,\Ohn, \Bon = 0\right) =  \tau_0\left(\frac{1}{\sqrt{1 - \left(\Ohn/\Ohc\left(\Wen\right)\right)^{2}}}\right),
\end{align}}

\noindent which is in quantitative agreement with our simulation data (figure~\ref{fig:Ohlim}\revRev{b}) when the critical Ohnesorge number \revRev{$\Ohc(\Wen)$} at which bouncing stops is taken from the simulations \revRev{(see the inset of figure \ref{fig:Ohlim}b)}. 
\citet{jha2020viscous} also predict the coefficient of restitution, written in our notations, as

\revRev{\begin{align}
	\label{eqn:JhaEtAl_epsilon}
	\varepsilon\left(\Wen, \Ohn, \Bon = 0\right) = \varepsilon_0\left(\Wen\right)\exp\left( \frac{-\beta \Ohn/\Ohc\left(\Wen\right)}{\sqrt{1 - \left(\Ohn/\Ohc\left(\Wen\right)\right)^{2}}} \right),
\end{align}}

\noindent where $\beta$ is an adjustable \revRev{$\Wen$-independent} parameter. The simulation data and the model are in excellent agreement for \revRev{$\beta = 4.00 \pm 0.25$} (figure~\ref{fig:Ohlim}\revRev{a}). 
Note that \citet{jha2020viscous} further reduced equation~\eqref{eqn:JhaEtAl_epsilon} to $\varepsilon\left(\Wen, \Ohn, \Bon = 0\right) \approx \varepsilon_0\left(\Wen\right)\exp\left(-\alpha\Ohn\right)$ for $\Ohn \ll \Ohc$, where $\alpha = \beta/\Ohc = 2.5 \pm 0.5$ best fits the experimental data, independent of the impact Weber number. 
The equivalent fitting parameter for our case is $\alpha^{\prime} = \beta^\prime/\Ohc = 3 \pm 1$, in very good agreement with the value reported by \citet{jha2020viscous}, despite the different Bond number \citep[$\Bon = 0$ here {\it vs.} $\Bon = 0.167$ for][also see \S~\ref{sec:LimitingCases:Bo}, and appendix~\ref{app:Weber}]{jha2020viscous}. 

Finally, we discuss the failure of the model of \citet{jha2020viscous} to predict the low $\Ohn$ behavior of the coefficient of restitution, that is contained in the prefactor $\varepsilon_0$.
The analysis of the overall energy \revRev{budget} shows that two ingredients are responsible for the loss of the initial drop kinetic energy in the $\Ohn < 0.01$ limit: (i) the presence of excess surface energy at take-off $\Delta E_\gamma(t_c)$ and (ii) the viscous dissipation in thin boundary layers at the liquid--air interface.
Both these contributions are not accounted for in the model of \citet{jha2020viscous} which takes no deformation, {\it i.e.}, $\Delta E_\gamma(t_c) = 0$ as a take-off condition, and $\eta_d V R$ as the scaling form of the viscous damping term, added to the liquid spring, which supposes that dissipation occurs at the drop length scale. 
However, as $\Ohn$ increases, bulk dissipation becomes dominant, explaining the ability of the model to capture bouncing inhibition.

\subsection{How does a heavy drop stop bouncing?}\label{sec:LimitingCases:Bo}

\begin{figure}
	\centering
	\includegraphics[width=\textwidth]{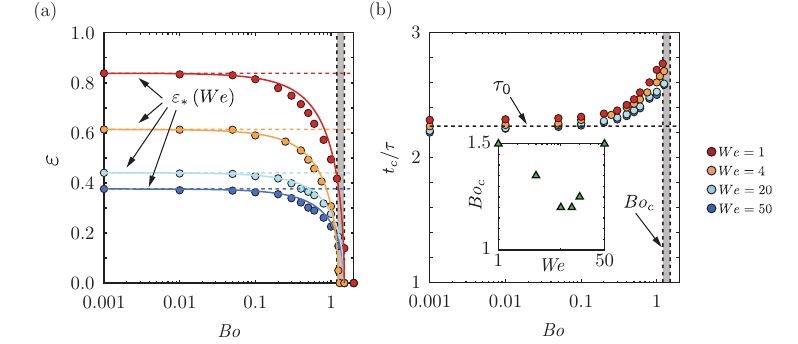}
	\caption{\revRev{Variation of (a) the restitution coefficient $\varepsilon$, and (b) the contact time $t_c$, normalized by the inertio-capillary timescale $\tau = \sqrt{\rho_dR^3/\gamma}$,} with the Bond number $\Bon$ at different Weber numbers $\Wen$ in the so-called inviscid regime ($\Ohn = 0.01$). In panel (b), the solid lines represent the predictions of the model of \citet{biance2006}, equation \eqref{eqn:BianceEtAl_epsilon}. The horizontal dashed lines represent the contact time and restitution coefficient values in the limit of zero Bond number ($\varepsilon\left(\Wen, \Ohn = 0.01, \Bon = 0\right)$), while the black vertical lines and the gray shaded regions mark the critical Bond number $\Boc \sim \mathcal{O}\left(1\right)$ beyond which drops do not bounce. \revRev{Note that the model of \citet{biance2006} predicts a constant $t_c$ (horizontal dashed line in panel a), in contradiction to our numerical simulations.}}
	\label{fig:Bolim}
\end{figure}
\begin{figure}
	\centering
	\includegraphics[width=\textwidth]{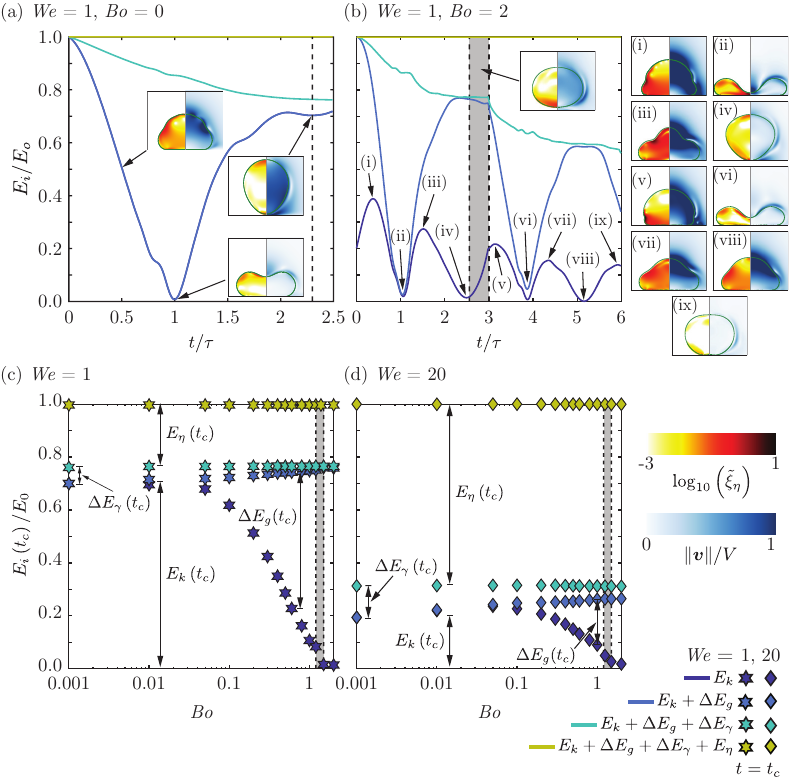}
	\caption{Energy budgets for drop impacts with $\Ohn = 0.01$ and $\Wen = 1$ for $\Bon = 0$ (a) and $\Bon = 2$ (b). $E_k$ and $E_\eta$ represent the kinetic energy and viscous dissipation, respectively. $\Delta E_g$ and $\Delta E_\gamma$ denote the the change in gravitational potential energy and surface energy with their zeroes set at the instant of maximum spreading of the impacting drop, and at $t = 0$, respectively. 
	The numerical snapshots in the insets illustrate the drop morphologies and the anatomy of the internal flow with a color code identical to that of figure \ref{fig:Phenomenology}.
	The vertical dotted line in panel (a) marks the instant when the drop takes off. In panel (b), the black vertical lines and the gray shaded regions bounds the time interval when the normal contact force between the drop and the substrate is zero. (c, d) Energy distributions at $t = t_c$ for $\Wen = 1$ (c) and $\Wen = 20$ (d) as a function of $\Bon$. For non-bouncing cases, $t_c$ represents the end of first drop oscillation cycle.
	The black vertical lines and the gray shaded regions in panels (c) and (d) mark the critical Bond number $\Boc \sim \mathcal{O}\left(1\right)$ beyond which drops do not bounce. See also supplementary movie~\red{SM4}.}
	\label{fig:BolimDescription}
\end{figure}

We now discuss the rebound inhibition of heavy drops, much larger than than their visco-capillary lengths, {\it i.e.}, with $\Ohn \ll 1$. We study this limit, in which equation \eqref{eq:MainEquation} reduces to $\Boc \sim 1$, by fixing $\Ohn = 0.01$, in the so-called inviscid bouncing regime (figures~\ref{fig:Ohlim} and~\ref{fig:OhlimDescription}), and by varying the Bond number $\Bon$. 

In figure~\ref{fig:Bolim}, we show the evolution of the coefficient of restitution $\varepsilon$ and of the normalized contact time $t_c/\tau$ as a function of the Bond number $\Bon$ for four values of the Weber number.
The variation of $\varepsilon$ with $Bo$ is qualitatively similar to that observed when sweeping across the viscous drop asymptote.
The coefficient of restitution $\varepsilon$ slowly decreases from its Weber--dependent value $\varepsilon_*(\Wen) = \varepsilon(\Wen,\Ohn = 0.01,\Bon = 0)$ with increasing Bond number, until it approaches a critical Bond number $\Boc$, of order one, at which it sharply decreases to zero.
Here, in the so-called inviscid regime, $\varepsilon_*(\Wen) \approx \varepsilon_0(\Wen)$. However, the influence of $\Bon$ and $\Ohn$ on $t_c$ are different. The contact time value hardly deviates from its inertio-capillary value, $\tau_0 = 2.25 \tau$, when varying $\Bon$ over two orders of magnitude. Yet, we only observe a moderate increase of $t_c$ as $\Boc$ is approached, contrasting with the divergence of $t_c$ close to $\Ohc$.

Figure~\ref{fig:Bolim} also evidences that varying the Weber number $\Wen$ from 1 to 50 hardly affects the critical Bond number $\Boc$ \revRev{(see the inset of figure~\ref{fig:Bolim}b)}, marking the transition from bouncing to floating, as underlined by the gray shaded regions and in agreement with the transition criterion \eqref{eq:MainEquation}. Similarly as for viscous drops, increasing $\Wen$ does not influence the contact time but markedly decreases $\varepsilon_*$, an effect we quantify in appendix \ref{app:Weber}.

To further investigate the variation of $\varepsilon$ with $\Bon$, we compute the overall energy budget during an impact event. In the presence of gravity, the energy balance \eqref{eqn:OhEnergyBalance} incorporates an additional contribution from the drop's gravitational potential energy, $\Delta E_g$, whose zero is set at the instant of maximum drop deformation. The modified energy balance reads

\begin{align}
	\label{eqn:BoEnergyBalance}
	\tilde{E}_0 = \tilde{E}_k(\tilde{t}) + \Delta\tilde{E}_g(\tilde{t}) + \Delta\tilde{E}_\gamma(\tilde{t}) + \tilde{E}_\eta(\tilde{t}),
\end{align}

\noindent where the initial energy also includes gravity, $\tilde{E}_0 = (4\pi/3)\left(\Wen/2 + \Bon(1-\mathcal{H})\right)$, with $\mathcal{H}$ denoting the \revRev{centre} of mass height of the drop at maximum deformation. 

Figure~\ref{fig:BolimDescription}(a) illustrates the energy budget for $(\Wen, \Ohn, \Bon) = (1, 0.01,0)$. The energy transfer follows similar dynamics as that described in figure~\ref{fig:OhlimDescription}(a) where $(\Wen, \Ohn, \Bon) = (1, 0.001, 0)$. The fraction of the initial kinetic energy recovered at take-off is the same, $E_k(t_c) \approx 0.75 E_0$. We also note that although $\Ohn$ has increased by an order of magnitude compared to the case shown in figure~\ref{fig:OhlimDescription}(a), the energy lost to viscous dissipation still accounts for a similar fraction of the initial energy, $E_\eta(t_c)\approx 0.2 E_0$, as expected in the so-called inviscid drop limit. The snapshots of the drop's internal flow (see the insets of figures \ref{fig:OhlimDescription}a and \ref{fig:BolimDescription}a) give insight into the independence of $E_\eta(t_c)$ with $\Ohn$. As the drop Ohnesorge number is increased, two antagonistic effects take place: (i) the viscous boundary layer grows larger, increasing dissipation, and (ii) capillary waves are attenuated, decreasing local dissipation. This competition qualitatively explains the independence of $E_\eta(t_c)$ and $\varepsilon$ on the drop Ohnesorge number for $\Ohn<0.01$.

Increasing the Bond number to $\Bon = 2$, beyond $\Boc$, sheds light on the mechanism of bouncing inhibition of heavy drops. At $t=0$, the drop has a higher initial energy owing to the contribution from the gravitational potential energy. As a result, the kinetic energy $E_k$ increases until the inertial shock is propagated throughout the drop \citep[see figure~\ref{fig:BolimDescription}b-i and][]{Gordillo2018, cheng2021drop}, before decreasing and reaching a minimum as the drop then attains maximum \revRev{deformation} (figure~\ref{fig:BolimDescription}b-ii).
In spite of these differences, the maximal spreading time is the same as that observed at low $\Bon$, $t_m \approx \tau$ \revRev{\citep[for $We = 1$, see][]{zhang2022impact}}, and viscous dissipation enervates a similar proportion of the initial energy as in the low $\Bon$ case during spreading.
During the retraction stage, $E_k$ increases (figure~\ref{fig:BolimDescription}b-ii to~\ref{fig:BolimDescription}b-iii), until the motion goes from being dominantly in the radial direction to being dominantly in the axial direction \citep[figure~\ref{fig:BolimDescription}b-iii, $t \approx 1.5\tau$, see][]{chantelot2018rebonds, zhang2022impact}.
Beyond this instant, gravity opposes the upward motion of the drop, $E_k$ decreases and is mainly transferred to $E_g$ until, eventually, at $t \approx 2.5\tau$ (figure~\ref{fig:BolimDescription}b-iv) the drop's \revRev{centre} of mass starts moving in the downward direction.
At this instant, only $20\%$ of the drop's initial energy goes to viscous dissipation, identical to the case of $\Bon = 0$, but bouncing is inhibited.
In contrast to the viscous asymptote, energy is still available to the drop even though the rebound is suppressed.
Subsequently, the drop undergoes several capillary oscillations at the substrate with a time period of approximately $2.5\tau$ (figure~\ref{fig:BolimDescription}b-v to~\ref{fig:BolimDescription}b-ix).

Figures \ref{fig:BolimDescription}(c) and (d) show the distribution of energy at take-off as a function of $\Bon$ for $\Wen = 1$ and $\Wen =20$, respectively.
For both Weber numbers, as $\Bon$ increases, the fraction of initial energy that goes into viscous dissipation, $E_\eta(t_c)$, is constant. 
However, the gravitational potential energy $\Delta E_g(t_c)$, initially negligible for $\Bon = 0.001$, increases with increasing $\Bon$, leading to a decrease of the drop's kinetic energy at take-off $E_k(t_c)$, which eventually drops to zero as bouncing stops at $\Boc$. 
\revRev{Noticing that energy is still available to the drop in the form of gravitational potential energy at $t=t_c$} allows us to rationalize the different behavior of $t_c$ with $\Bon$ and $\Ohn$. Indeed, the viscous rebound suppression corresponds to a transition from an underdamped to overdamped system, associated to a divergence of the oscillation period, while heavy drops undergo successive energy transfers between gravitational potential, kinetic and surface energy on the inertio-capillary timescale.

Finally, we compare the variation of $\varepsilon$ and $t_c$ with $\Bon$ extracted from our simulations to the spring-mass model of \citet{biance2006} which takes into account the role of gravity but neglects viscous dissipation.
In our notation, the dependence of $\varepsilon$ on $\Bon$ \revRev{in the model of \citet{biance2006}} is written as

\revRev{\begin{align}
	\label{eqn:BianceEtAl_epsilon} 
	\varepsilon\left(\Wen, \Ohn = 0.01, \Bon\right) = \varepsilon_*\left(\Wen\right)\sqrt{\left(1-\Bon/\Boc\left(\Wen\right)\right)\left(1 + \Bon/\left(3\Boc\left(\Wen\right)\right)\right)}.
\end{align}}

\noindent \revRev{Equation~\eqref{eqn:BianceEtAl_epsilon} is in excellent quantitative agreement with the values of \revRev{$\varepsilon_*(\Wen)$} and \revRev{$\Boc(\Wen)$} extracted from the simulations (figure \ref{fig:Bolim}a). However, this model predicts a constant $t_c$ which is in disagreement with our simulations as we approach $Bo_c$ (figure \ref{fig:Bolim}b)}. We further stress that the model of \citet{biance2006}, similarly as that of \citet{jha2020viscous}, does not capture the evolution of the prefactor $\varepsilon_*$ or $\varepsilon_0$ with $\Wen$. It indeed does not take into account viscous dissipation, which we have shown to be the main ingredient responsible for the loss of rebound elasticity as $\Wen$ is increased. We address the relevance of this model in predicting the variation of the coefficient of restitution with $\Wen$ in appendix~\ref{app:Weber}.

\section{Conclusion and outlook}\label{sec:Conclusion}

Drops smaller than their visco-capillary length, {\it i.e.}, with $\Ohn > 1$, stop bouncing due to viscous dissipation, while those larger than their gravito-capillary length, {\it i.e.}, with $\Bon > 1$, cannot bounce due to their weight. 
In this \revRev{paper}, we investigate how viscous stresses and gravity 
\revRev{oppose} capillarity to prevent drops of intermediate sizes, with $\eta_d^2/\rho_d\gamma < R < \sqrt{\gamma/\rho_d g}$, \revRev{corresponding to $\Bon < 1$ and $\Ohn < 1$}, from bouncing off non-wetting substrates.
We emphasize the relevance of this regime, which describes the bouncing inhibition of millimetre--sized aqueous or silicone oil drops, commonly used in experiments. Drawing an analogy with coalescence-induced jumping of two identical drops, we propose the criterion $\Ohc + \Boc \sim 1$ for the bouncing to non-bouncing transition. Through a series of direct numerical simulations, we show the validity of this criterion over a wide range of Weber numbers in the $\Bon$--$\Ohn$ phase space.

We also study the details of the mechanism of rebound suppression in the two limiting cases of low drop Ohnesorge number and Bond number, by relating \revRev{the} overall energy \revRev{budget} to the drop morphology and flow anatomy. For drops much smaller than their gravito-capillary lengths ($\Bon \ll 1$), simulations confirm that the increase of viscous dissipation in the bulk of the liquid is responsible for rebound suppression. The restitution coefficient decays exponentially with increasing $\Ohn$ until a critical Ohnesorge number $\Ohc$, of order one, is reached and the contact time diverges as the rebound process becomes over--damped.
This behaviour is well captured by the model of \citet{jha2020viscous} which extends the spring--mass analogy to viscous drops by including bulk viscous dissipation in the liquid. 
Moreover, the energy budgets reveal that the rebound elasticity in the so-called inviscid regime $\Ohn < 0.01$, in which the restitution coefficient is independent of $\Ohn$, is controlled by viscous dissipation occuring in thin boundary layers, shedding light on the failure of simple scaling models to capture this regime. 
We also evidence that the decrease of rebound elasticity with increasing $\Wen$ in the so-called inviscid regime is a consequence of enhanced viscous dissipation, as the surface energy stored at take-off plays a negligible role in setting $\varepsilon$.

For drops much larger than their visco-capillary lengths ($\Ohn \ll 1$), the excess gravitational potential energy at take-off stops \revRev{the drop from rebounding} when the Bond number reaches a critical value $\Boc$, of order one.
Indeed, an increase in $\Bon$ does not change the fraction of the drop initial energy that goes into viscous dissipation during the impact process.
The restitution coefficient deviates slowly from its $\Wen$-dependent value at zero Bond number, until it decreases sharply to zero as $\Boc$ is reached.
This decrease is quantitatively captured by the spring--mass model of \citet{biance2006}, which takes into account the effect of gravity.
We stress that, as \revRev{the} rebound is suppressed, energy is still available to the drop which subsequently oscillates on the substrate on the inertio-capillary timescale. Contrary to viscous bouncing inhibition, the rebound suppression of heavy drops is not associated to a divergence of the contact time.

Finally, we emphasize that this work describes the upper bound of the bouncing to non-bouncing transition on ideal non-wetting substrates. Indeed, water drops can cease bouncing due to substrate pinning on superhydrophobic substrates \citep{sarma2022interfacial}. We also idealized the role of the surrounding medium by keeping a small value for the Ohnesorge number, $\Oha = 10^{-5}$. We anticipate that dissipation in the surrounding medium might play a role in the impact of microdrops as $\Oha$ increases \citep{kolinski2014drops, tai2021research}. Lastly, the influence of the Weber number on the elasticity of the impact process deserves further investigation. Here, we only focus on impacts with $\Wen \ge 1$, where the bouncing inhibition and drop contact time are reasonably insensitive to an increase in Weber number. Yet, modelling the full Weber number dependence of the restitution coefficient at both low $\Ohn$ and low $\Bon$ still demands further work. It will be particularly interesting to study the regime $\Wen \ll 1$, where drops only deform weakly, and the internal flow is still significant at the instant of maximum spreading.\\

\noindent{\bf  Supplementary data\bf{.}} \label{SM} Supplementary material and movies are available at xxxx \\

\noindent{\bf Acknowledgments\bf{.}} We thank Aditya Jha for sharing data and for stimulating discussions. We also thank Uddalok Sen, Maziyar Jalaal, Andrea Prosperetti, David Qu{\'e}r{\'e}, and Jacco Snoeijer for discussions. We acknowledge Srinath Lakshman for preliminary experiments that made us numerically and theoretically explore the effect of gravity in viscous drop bouncing. This work was carried out on the national e-infrastructure of SURFsara, a subsidiary of SURF cooperation, the collaborative ICT organization for Dutch education and research.\\

\noindent{\bf Funding\bf{.}} The authors acknowledge the ERC Advanced Grant No. 740479-DDD.\\

\noindent{\bf Declaration of Interests\bf{.}} The  authors report no conflict of interest. \\

\noindent{\bf Authors' ORCID\bf{.}} \\
V. Sanjay \href{https://orcid.org/0000-0002-4293-6099}{orcid.org/0000-0002-4293-6099}\\
P. Chantelot \href{https://orcid.org/0000-0003-1342-2539}{orcid.org/0000-0003-1342-2539}\\
D. Lohse \href{https://orcid.org/0000-0003-4138-2255}{orcid.org/0000-0003-4138-2255}.\\

\appendix

\section{Measuring the restitution coefficient}\label{app:restitution in simulations}

\begin{figure}
	\centering
	\includegraphics[width=\textwidth]{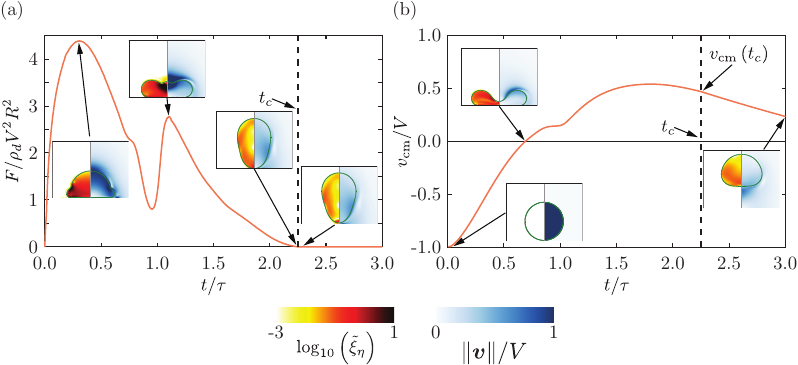}
	\caption{A representative temporal variation of (a) the normal reaction force $F$ on the drop and (b) its \revRev{centre} of mass velocity $v_{\text{cm}}$. Time is normalized using the inertio-capillary timescale $\tau$. Insets illustrate the different stages of drop impact process. The background shows the magnitude of the rate of viscous dissipation per unit volume ($\tilde{\xi}_\eta = 2Oh\left(\boldsymbol{\tilde{\mathcal{D}}:\tilde{\mathcal{D}}}\right)$) on the left and the magnitude of velocity field normalized by the impact velocity on the right. The vertical dashed black line represents the contact time calculated using the criterion, $F = 0$ marking the end of contact between the drop and the substrate. Here, $\left(\Wen, \Ohn, \Bon\right) = \left(4, 0.034, 0.5\right)$, the contact time $t_c = 2.25\tau$, and the coefficient of restitution $\varepsilon = 0.47$.}
	\label{fig:AppendixRestitution}
\end{figure}

Throughout this manuscript, we have used the time of contact and restitution coefficient to study the drop impact dynamics. In this appendix, we describe the procedure used to determine the restitution coefficient which is the ratio of take-off velocity $v_{\text{cm}}(t_c)$ to the impact velocity $V$,

\begin{align}
	\varepsilon = \frac{v_{\text{cm}}(t_c)}{V}
\end{align}

\noindent where $t_c$ denotes the contact time when the drop leaves the substrate. We assume an ideal non-wetting substrate by ensuring that a thin air layer (with a minimum thickness of $\Delta = R/1024$, where $\Delta$ is the minimum grid size employed in the simulations) is always present between the drop and the substrate \citep[also see][]{ramirezsoto-2020-sciadv}. Hence, we need to define a criterion for the end of contact. We do so at the instant when the normal reaction force $F$ between the substrate and the drop is zero \citep[for calculation details, see][]{zhang2022impact}, as shown in figure~\ref{fig:AppendixRestitution}(a). Subsequently, we read out the \revRev{centre} of mass velocity (figure~\ref{fig:AppendixRestitution}b) at this instant. If this \revRev{centre} of mass velocity is not in the upward direction ({\it i.e.}, it is zero or negative), we categorize the case as non-bouncing. For the representative case in figure~\ref{fig:AppendixRestitution}, $\varepsilon = 0.47$. 

\section{Influence of Weber number}\label{app:Weber}

\begin{figure}
	\centering
	\includegraphics[width=\textwidth]{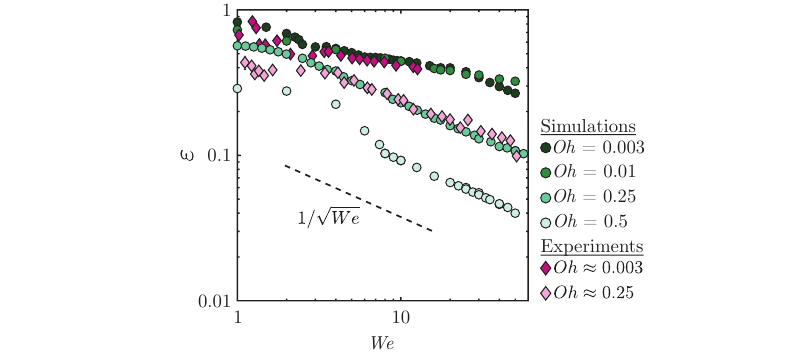}
	\caption{Variation of restitution coefficient with the impact Weber number ($\Wen$) at different drop Ohnesorge number ($\Ohn$). The simulations (circle data points) match perfectly with the experimental results (diamond data points) of \citet{jha2020viscous} without any fitting parameters. 
	Here, the Bond number ($\Bon$) is $0.167$.}
	\label{fig:AppendixWeber}
\end{figure}

We report that the bouncing inhibition and drop contact time are fairly insensitive to an increase in the impact Weber number ($\Wen$) while the restitution coefficient decreases monotonically with $\Wen$.
Figure~\ref{fig:AppendixWeber} illustrates the variation of the restitution coefficient with $\Wen$ at different $\Ohn$ and fixed $\Bon = 0.167$, enabling comparison with the experimental data of \citet{jha2020viscous}. 
In the so-called inviscid drop limit ($\Ohn \lesssim 0.01$), the coefficient of restitution is approximately equal to the prefactors $\varepsilon_0\left(\Wen\right)$ and $\varepsilon_*\left(\Wen\right)$ used in \S~\ref{sec:LimitingCases}, equations~\eqref{eqn:JhaEtAl_epsilon}, and~\eqref{eqn:BianceEtAl_epsilon}, respectively.

For $\Ohn \lesssim 0.01$, $\varepsilon$ does not follow the $1/\sqrt{\Wen}$ scaling relation derived by \citet{biance2006} using a spring-mass model that neglects the influence of the drop viscosity. 
Indeed, the energy budgets reported in figures \ref{fig:OhlimDescription}(c,d) and \ref{fig:BolimDescription}(c,d) evidence that the transfer to surface energy at takeoff $\Delta E_\gamma(t_c)$,  proposed by \citet{biance2006} to account for the loss of rebound elasticity, negligibly contribute to the decrease of $\varepsilon$ with $\Wen$. 
Instead, we find that the increase of viscous dissipation with $\Wen$ drives the decrease of $\varepsilon$, even in the so-called inviscid limit.

Interestingly, the restitution coefficient for viscous drop impacts ($\Ohn \gtrsim 0.1$) seems to follow the $1/\sqrt{\Wen}$ scaling relation, implying that the take-off velocity scales with the Taylor-Culick velocity ($v_{\text{cm}}(t_c) \sim \sqrt{\gamma/\rho_dR}$), and is independent of the impact velocity $V$, consistent with our assumption that the retraction and take-off stages are independent of the impact Weber number. 
We caution here that the range of $\Wen$ $\left(1 \le \Wen \le 50\right)$ is too small to claim these scaling relations convincingly. 

Lastly, notice the remarkable agreement between our simulations and the experimental data points from \citet{jha2020viscous} for two different drop Ohnesorge numbers, which differ by over two orders of magnitude (see figure \ref{fig:AppendixWeber}).


\section{Code availability}
The codes used in the present article are permanently available at \citet{basiliskVatsalViscousBouncing}.

\bibliographystyle{jfm}
\bibliography{ViscousDropImpact}

\end{document}